\documentclass[pre,twocolumn,superscriptaddress,preprintnumbers,floatfix]{revtex4}

\usepackage{dcolumn}
\usepackage{amsmath}
\usepackage{amssymb}
\usepackage{graphicx}
\usepackage{bm}   % bold math
\usepackage{bbm}   
\usepackage{verbatim}
\usepackage{stmaryrd}
\usepackage{amsthm}

\theoremstyle{break} 	
\theoremstyle{plain} 	
\theoremstyle{break} 	
\theoremstyle{plain} 	
\theoremstyle{break} 	
\theoremstyle{plain} 	
\theoremstyle{plain}	
\theoremstyle{break}	 
\theoremstyle{plain}	
\theoremstyle{break}	 

\def\clap#1{\hbox to 0pt{\hss#1\hss}}

\def\mathclap{\mathpalette\mathclapinternal}

\def\mathclapinternal#1#2{%
  \clap{$\mathsurround=0pt#1{#2}$}}
\def\bra#1{\mathinner{\langle{#1}|}}
\def\ket#1{\mathinner{|{#1}\rangle}}
\def\braket#1{\mathinner{\langle{#1}\rangle}}

\newcommand{\Prob}      {\mathrm{Pr}}

\newcommand{\stochL}    {\mathcal{L}}

\newcommand{\svector}[1]{\bra{ {#1} }}
\newcommand{\past}{\stackrel{\leftarrow}{S}}

\newcommand{\future}{\stackrel{\rightarrow}{S}}

\def\Abet {{\mathcal A}}
\def\EE   {{\bf E} }
\def\symb {s}
\def\hmu   {h_\mu} 
\def\l2   {{\rm log}_2}

\begin{document}

\title{Computation in Sofic Quantum Dynamical Systems}

\author{Karoline Wiesner}
\email{karoline@cse.ucdavis.edu}
\affiliation{Center for Computational Science \& Engineering and Physics Department,
University of California Davis, One Shields Avenue, Davis, CA 95616}
\author{James P. Crutchfield}
\email{chaos@cse.ucdavis.edu}
\affiliation{Center for Computational Science \& Engineering and Physics Department,
University of California Davis, One Shields Avenue, Davis, CA 95616}

\bibliographystyle{unsrt}

\begin{abstract}
We analyze how measured quantum dynamical systems store and process
information, introducing sofic quantum dynamical systems. Using recently
introduced information-theoretic measures for quantum processes, we quantify
their information storage and processing in terms of \emph{entropy rate} and
\emph{excess entropy}, giving closed-form expressions where possible. To
illustrate the impact of measurement on information storage in quantum
processes, we analyze two spin-$1$ sofic quantum systems that differ only
in how they are measured.
\end{abstract}

\maketitle

\section{Introduction}

Extending concepts from symbolic dynamics to the quantum setting, we forge a
link between quantum dynamical systems and quantum computation, in general,
and with quantum automata, in particular.

Symbolic dynamics originated as a method to study general dynamical systems
when, nearly 100 years ago, Hadamard used infinite sequences of symbols to
analyze the structure of geodesics on manifolds of negative curvature; see
Ref.~\cite{kitchens} and references therein. In the 1930's and 40's Hedlund
and Morse coined the term \emph{symbolic dynamics} \cite{Hedl38a,Hedl40a} to
describe the study of dynamics over the space of symbol sequences in their
own right. In the 1940's Shannon used sequence spaces to describe information
channels \cite{Shan62}. Subsequently, the techniques and ideas have found
significant applications beyond dynamical systems, in data storage and
transmission, as well as in linear algebra \cite{lind}. 

On the flip side of the same coin, computation theory codes symbol sequences
using finite-state automata. The class of sequences that can be coded this way
define the \emph{regular languages} \cite{hopcroft}. It turns out that many
dynamical systems can also be coded with finite-state automata using the tools
of symbolic dynamics \cite{lind}. In fact, \emph{Sofic systems} are the
particular class of dynamical systems that are the analogs of regular
languages in automata theory.

The study of quantum behavior in classically chaotic systems is yet another
active thread in dynamical systems \cite{gutzwiller, reic04a} which has most
recently come to address the role of measurement. Measurement interaction
leads to genuinely chaotic behavior in quantum systems, even far from the
semi-classical limit \cite{habi06}. Classical dynamical systems can be
embedded in quantum dynamical systems as the special class of commutative
dynamical systems, which allow for unambiguous assignment of joint
probabilities to two observations \cite{alicki}.

An attempt to construct symbolic dynamics for quantum dynamical systems was
made by Alicki and Fannes \cite{alicki}, who defined shifts on spin chains as
the analog of shifts in sequence space. Definitions of entropy followed from
this. However, no connection to quantum finite-state automata was established
there.

In the following we develop an approach that differs from this and other
previous attempts since it explicitly accounts for observed sequences, in
contrast to sequences of
(unobservable) quantum objects, such as spin chains. The recently introduced
computational model class of \emph{quantum finite-state generators} provides
the required link between quantum dynamical systems and the theory of automata
and formal languages \cite{wies06b}. It gives access to an analysis of quantum
dynamical systems in terms of symbolic dynamics. Here, we strengthen that link
by studying explicit examples of quantum dynamical systems. We construct their
quantum finite-state generators and establish their \emph{sofic} nature. In
addition we review tools that give an information-theoretic analysis for
quantifying the information storage and processing of these systems. It turns
out that both the sofic nature and information processing capacity depend on
the way a quantum system is measured.

%%%%%%%%%%%%%%%%%%%%%%%%%%%%%%%%%%%%%%%%%%%%%%%%%%%%%%%%%%%%%%%%%%%%%%%

\section{Quantum finite-state generators}

To start, we recall the \emph{quantum finite-state generators} (QFGs)
defined in Ref. \cite{wies06b}. They consist of a finite set of
\emph{internal states} $Q = \{q_i: i = 1, \ldots, |Q| \}$. The
\emph{state vector} is an element of a $|Q|$-dimensional Hilbert
space: $\bra{\psi} \in \mathcal{H}$. At each time step a
quantum generator outputs a symbol $s \in \Abet$ and updates its
state vector.

The temporal dynamics is governed by a set of $|Q|$-dimensional
\emph{transition matrices} $\{T(s) = U \cdot P(s), s \in \Abet \}$,
whose components are elements of the complex unit disk and where each
is a product of a unitary matrix $U$ and a projection operator $P(s)$.
$U$ is a $|Q|$-dimensional unitary \emph{evolution operator} that
governs the evolution of the state vector $\bra{\psi}$.
$\mathbf{P} =\{ P(s): s \in \Abet \}$
is a set of \emph{projection operators}---$|Q|$-dimensional Hermitian
matrices---that determines how the state vector is measured. The operators
span the Hilbert space: $\sum_\symb P(s) = \mathbbm{1}$.

Each output symbol $s$ is identified with the measurement outcome and labels
one of the system's eigenvalues. The projection operators determine how output
symbols are generated from the internal, hidden unitary dynamics. They are
the only way to observe a quantum process's current internal state. 

A quantum generator operates as follows. $U_{ij}$ gives the transition
amplitude from internal state $q_i$ to internal state $q_j$. Starting in
state vector $\bra{\psi_0}$ the generator updates its state by applying the
unitary matrix $U$. Then the state vector is projected using $P(s)$ and
renormalized. Finally, symbol $s \in \Abet$ is emitted. In other words,
starting with state vector $\svector{\psi_0}$, a single time-step yields
$\bra{ \psi(s) } = \bra{\psi_0} U \cdot P(s)$, with the observer
receiving measurement outcome $s$.

% **********************************************************************
\subsection{Process languages}

The only physically consistent way to describe a quantum system under iterated
observation is in terms of the observed sequence
$\stackrel{\leftrightarrow}{S} \, \equiv \,\ldots S_{-2} S_{-1} S_0 S_1 \ldots$
of discrete random variables $S_t$. We consider the family of
\emph{word distributions}, 
$\{ {\rm Pr}(s_{t+1} , \ldots , s_{t+L}): s_t \in \Abet \}$, where
${\rm Pr}(s_t)$ denotes the probability that at time $t$ the random
variable $S_t$ takes on the particular value $s_t \in \Abet$ and
${\rm Pr} (s_{t+1} , \ldots , s_{t+L})$ denotes the joint probability over
sequences of $L$ consecutive measurement outcomes. We assume that the
distribution is stationary:
\begin{equation}
{\rm Pr}(S_{t+1},\ldots, S_{t+L})={\rm Pr}(S_1, \ldots , S_L ) ~. 
\end{equation}
We denote a block of $L$ consecutive variables by $S^L \equiv S_1 \ldots S_L$
and the lowercase $s^L = s_1 s_2 \cdots s_{L}$ denotes a particular
measurement sequence of length $L$. We use the term {\em quantum process}
to refer to the joint distribution
${\rm Pr} (\stackrel{\leftrightarrow}{S})$ over the infinite chain of random
variables. A quantum process, defined in this way, is the quantum analog of
what Shannon referred to as an \emph{information source} \cite{cover}.

Such a quantum process can be described as a \emph{stochastic language}
$\stochL$, which
is a \emph{formal language} with a probability assigned to each word. A
stochastic language's word distribution is normalized at each word length:
\begin{align}
\sum_{\mathclap{\{\symb^L \in \stochL\}}} \Prob(\symb^L) = 1 ~, L = 1, 2, 3, \ldots
\end{align}
with $0 \leq \Prob(\symb^L) \leq 1$ and the consistency condition
$\Prob(\symb^L) \leq \Prob(\symb^L\symb)$.

A \emph{process language} is a stochastic language that is
\emph{subword closed}: all subwords of a word are in the language.

We can now determine word probabilities produced by a QFG. Starting the
generator in $\bra{\psi_0}$, the probability of output symbol $s$ is
given by the state vector without renormalization:
\begin{equation}
\label{eqn:qpry}
\Prob(s) =  \braket{ \psi(s) | \psi(s)} ~.
\end{equation}
While the probability of outcome $s^L$ from a measurement sequence is
\begin{equation}
\label{eqn:qprsL}
\Prob(s^L) =  \braket{ \psi(s^L) | \psi(s^L)} ~.
\end{equation}

In \cite{wies06b} the authors established a hierarchy of process languages and
the corresponding quantum and classical computation-theoretic models that can
recognize and generate them.

% **********************************************************************
\subsection{Alternative quantum finite-state machines}

Said most prosaically, we view quantum generators as representations of the
word distributions of quantum process languages. Despite similarities, this
is a rather different emphasis than that used before. The first mention of
\emph{quantum automata} as an empirical description of physical properties
was made by Albert in 1983 \cite{albe83}. Albert's results were subsequently
criticized by Peres for using an inadequate notion of measurement \cite{pere84}.
In a computation-theoretic context, quantum finite automata were introduced by
several authors and in varying ways, but all as devices for recognizing
word membership in a language. For the most widely discussed quantum automata,
see Refs.~\cite{moor00,kond97,ahar98}. Ref.~\cite{amba02} summarizes the
different classes of languages which they can recognize. Quantum transducers
were introduced by Freivalds and Winter \cite{frei01}. Their definition,
however, lacks a physical notion of measurement. We, then, introduced quantum
finite-state machines, as a type of transducer, as the general object that
can be reduced to the special cases of quantum recognizers and quantum
generators of process languages \cite{wies06b}.

% **********************************************************************
\section{Information processing in a spin-$1$ dynamical system}

We will now investigate concrete examples of quantum processes.
Consider a spin-$1$ particle subject to a magnetic field
which rotates the spin. The state evolution can be described by the
following unitary matrix:
\begin{equation}
U = \left(
	\begin{array}{ccc}
		\frac{1}{\sqrt{2}} & \frac{1}{\sqrt{2}} & 0
		\\ 0 & 0 & -1
		\\ -\frac{1}{\sqrt{2}} & \frac{1}{\sqrt{2}} & 0 
	\end{array}
	\right) ~.\\
\label{eqn:qgldm}
\end{equation}
Since all entries are real, $U$ defines a rotation in $\mathbb{R}^3$ around the y-axis by
angle $\frac{\pi}{4}$ followed by a rotation around the x-axis by
an angle $\frac{\pi}{2}$.

Using a suitable representation of the spin operators $J_i$
\cite[p. 199]{peres}:
\begin{align}
J_x & = \left(
    \begin{array}{ccc}
        0 & 0 & 0 \\
        0 & 0 & i \\
        0 & -i & 0
    \end{array}
    \right) ,~
J_y = \left(
    \begin{array}{ccc}
        0 & 0 & i \\
        0 & 0 & 0 \\
        -i & 0 & 0
    \end{array}
    \right) ,~ \nonumber \\
J_z & = \left(
    \begin{array}{ccc}
        0 & i & 0 \\
        -i & 0 & 0 \\
        0 & 0 & 0
    \end{array}
    \right) ,
\end{align}
the relation $P_i = 1 - J_i^2$
defines a one-to-one correspondence between the projector $P_i$ and the
square of the spin component along the $i$-axis. The resulting measurement
answers the yes-no question, Is the square of the spin component along
the $i$-axis zero?

Consider the observable $J_y^2$. Then the following projection operators
together with $U$ in Eq.~(\ref{eqn:qgldm}) define the quantum finite-state
generator:
\begin{align}
P(0) & = \ket{010}\bra{010} \nonumber \\
 ~\mathrm{and}~
~P(1) & = \ket{100}\bra{100} + \ket{001}\bra{001}
 ~.
\end{align}

\begin{figure}
\begin{center}
\resizebox{1.80in}{!}{\includegraphics{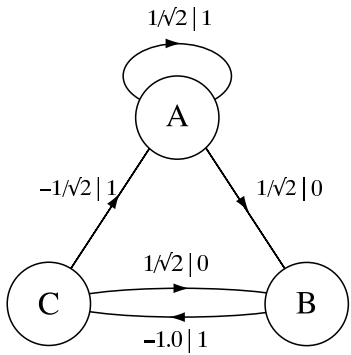}}
\end{center}
\caption{The Golden Mean quantum generator.
  }
\label{fig:gm-qdg}
\end{figure}

A graphical representation of the automaton is shown in Fig.~\ref{fig:gm-qdg}. 

\begin{figure}
\begin{center}
\resizebox{!}{2.60in}{\includegraphics{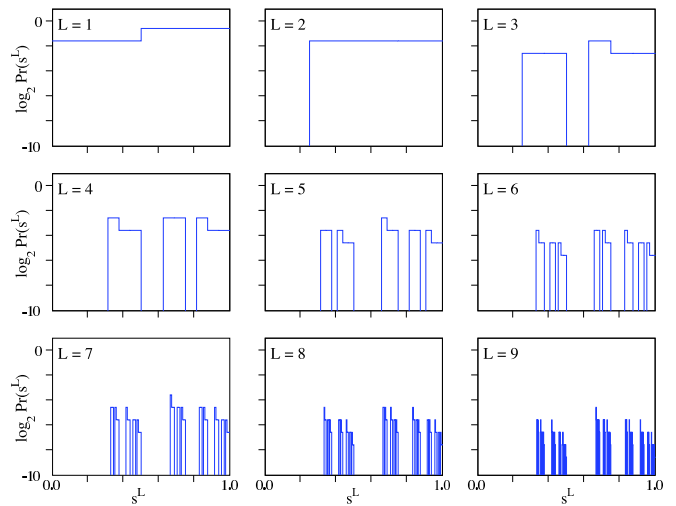}}
\end{center}
\caption{Golden Mean process language: Word $\{00\}$ 
  has zero probability; all others have nonzero probability. The logarithm
  base 2 of the word probabilities is plotted
  versus the binary string $\symb^L$, represented as base-$2$ real number
  ``$0.\symb^L$''. To allow word probabilities to be compared at different
  lengths, the distribution is normalized on $[0,1]$---that is, the
  probabilities are calculated as densities.
  }
\label{fig:gm-lang}
\end{figure}

The process language generated by this QFG is the so-called
\emph{Golden-Mean Process} language \cite{kitchens}. The word
distribution is shown in Fig.~\ref{fig:gm-lang}. It is
characterized by the set of \emph{irreducible forbidden words}
$\mathcal{F} = \{00\}$: no consecutive zeros occur. In other words,
for the spin-$1$ particle the spin component along the
$y$-axis never vanishes twice in a row. This restriction---the
dominant structure in the process---is a 
\emph{short-range correlation} since the 
measurement outcome at time $t$ only depends on the immediately preceding one
at time $t-1$. If the outcome is $0$, the next outcome will be
$1$ with certainty. If the outcome is $1$, the next measurement is
maximally uncertain: outcomes $0$ and $1$ occur with equal probability.

\begin{figure}
\begin{center}
\resizebox{1.80in}{!}{\includegraphics{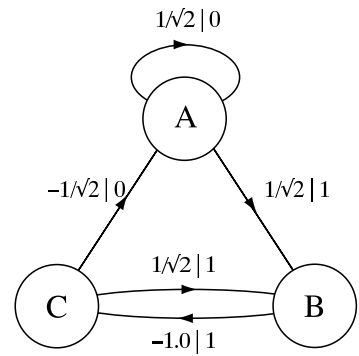}}
\end{center}
\caption{The Even Process quantum generator.
  }
\label{fig:ep-qdg}
\end{figure}

\begin{figure}
\begin{center}
\resizebox{!}{2.60in}{\includegraphics{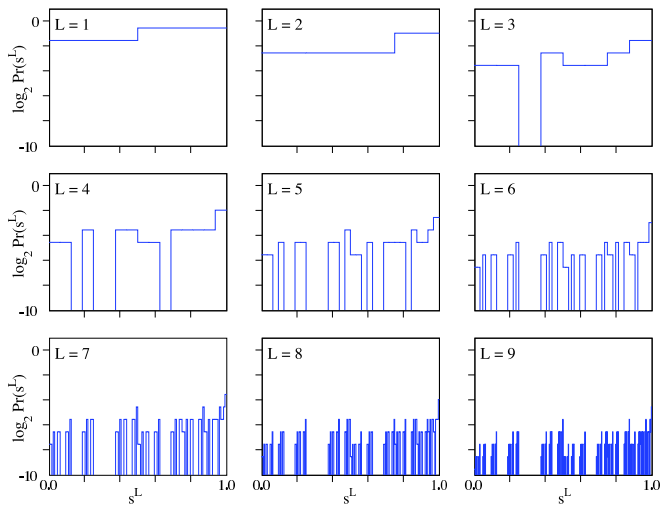}}
\end{center}
\caption{Even Process language: Words $\{01^{2k-1}0\}, k=1,2,3,...$ 
  have zero probability; all others have nonzero probability.
  }
\label{fig:ep-lang}
\end{figure}  

Consider the same Hamiltonian, but now use instead the observable $J_x^2$.
The corresponding projection operators define the QFG:
\begin{align}
\notag
P(0) &= \ket{100}\bra{100}\\
 ~\mathrm{and}~
~P(1) &= \ket{010}\bra{010} + \ket{001}\bra{001}
 ~.
\end{align}
The QFG defined by $U$ and these projection operators is shown in
Fig.~\ref{fig:ep-qdg}.
The process language generated by this QFG is the so-called
\emph{Even Process} language \cite{hirs70,kitchens}. The word distribution
is shown in Fig.~\ref{fig:ep-lang}. It is defined by the infinite set
of irreducible forbidden words $\mathcal{F} = \{01^{2k-1}0\}, k=1,2,3,...$.
That is, if the spin component equals 0 along the $x$-axis it will be zero
an even number of consecutive measurements before being observed to be
nonzero. This is a type of \emph{infinite correlation}: For a possibly
infinite number of time steps the system tracks the evenness or oddness
of number of consecutive measurements of ``spin component equals 0 along
the $x$-axis''.

Note that changing the measurement, specifically choosing $J_z^2$ as the
observable, yields a QFG that generates \emph{Golden Mean} process language
again.

The two processes produced by these quantum dynamical systems are well known
in the context of symbolic dynamics \cite{lind}---a connection we will return
to shortly. Let us first, though, turn to another important property of
finite-state machines and explore its role in computational capacity and
dynamics.

% **********************************************************************
\section{Determinism}

The label \emph{determinism} is used in a variety of senses, some of which are
seemingly contradictory. Here, we adopt the notion, familiar from
automata theory \cite{hopcroft}, which differs from that in physics, say, of
non-stochasticity. One calls a finite-state machine (classical or
quantum) \emph{deterministic} whenever the transition from one state to the
next is uniquely determined by the output symbol, or input symbol for
recognizers. It is important to realize that a \emph{deterministic}
finite-state machine can still behave stochastically---stochasticity here
referring to the positive probability of generating symbols. Once the symbol is
determined, though, the transition taken by the machine to the next state is 
unique. Thus, what is called a \emph{stochastic process} in dynamical systems
theory can be described by a deterministic finite-state generator without
contradiction.

We can easily check the two quantum finite-state machines in
Figs.~\ref{fig:gm-qdg} and \ref{fig:ep-qdg} for determinism by inspecting each
state and its outgoing transitions. One quickly sees that both generators are
deterministic. In contrast, the third QFG mentioned above, defined by $U$ in
Eq. (\ref{eqn:qgldm}) and $J_z^2$, is nondeterministic.

Determinism is a desirable property for various reasons. One is the simplicity
of the mapping between observed symbols and internal states. Once the observer
synchronizes to the internal state dynamics, the output symbols map one-to-one
\emph{onto} the internal states. In general, though, the observed symbol
sequences do not track the internal state dynamics (orbit) directly. (This
brings one to the topic of hidden Markov chains \cite{bishop}.)

For optimal prediction, however, access to the internal state dynamics is key.
Thus, when one has a deterministic model, the observed sequences reveal the
internal dynamics. Once they are known and one is synchronized, the process
becomes optimally predictable. A final, related reason why determinism is
desirable is that closed-form expressions can be given for various information
processing measures, as we will discuss in Sec.~\ref{sec:info}.

% **********************************************************************
\section{Sofic systems}
\label{sec:sofic}

In symbolic dynamics, sofic systems are used as tractable representations with
which to analyze continuous-state dynamical systems \cite{lind,kitchens}.
Let the alphabet $\Abet$ together with an $n \times n$ adjacency matrix (with
entries $0$ or $1$) define a directed graph $G=(V,E)$ with $V$ the set of
vertices and $E$ the set of edges. Let $X$ be the set of all infinite admissible
sequences of edges, where \emph{admissible} means that the sequence corresponds
to a path through the graph. Let $T$ be the shift operator on this sequence;
it plays the role of the time-evolution operator of the dynamical system. A
\emph{sofic system} is then defined as the pair $(X, T)$ \cite{weis73}. The
Golden Mean and the Even process are standard examples of \emph{sofic systems}.
The Even system, in particular, was introduced by Hirsch et al in the
1970s \cite{hirs70}.

Whenever the rule set for admissible sequences is finite one speaks of a
\emph{subshift of finite type}. The Golden Mean process is a subshift of
finite type. Words in the language are defined by the finite (single) rule
of not containing the subword $00$. The Even Process, on the other hand,
is not of finite type, since the number of rules is infinite: The forbidden
words $\{01^{2k+1}0\}$ cannot be reduced to a finite set. As we noted, the
rule set, which determines allowable words, implies the process has a kind
of infinite memory. One refers, in this case, to a \emph{strictly sofic}
system.

The spin-$1$ example above appears to be the first time a strictly sofic
system has been identified in quantum dynamics. This ties quantum dynamics
to languages and quantum automata theory in a way similar to that found in
classical dynamical systems theory. In the latter setting, words in the
sequences generated by sofic systems correspond to \emph{regular
languages}---languages recognized by some finite-state machine. We now have a
similar construction for
quantum dynamics. For any (finite-dimensional) quantum dynamical system under
observation we can construct a QFG, using a unitary operator and a set of
projection operators. The language it generates can then be analyzed in terms
of the rule set of admissible sequences. One interesting open problem becomes
the question whether the words produced by sofic quantum dynamical systems
correspond to the regular languages. An indication that this is not so is
given by the fact that finite-state quantum recognizers can accept nonregular
process languages \cite{wies06b}.

% **********************************************************************
\section{Information-theoretic analysis}
\label{sec:info}

The process languages generated by  the spin-$1$ particle under a particular
observation scheme can be analyzed using well known information-theoretic
quantities such as \emph{Shannon block entropy} and \emph{entropy rate}
\cite{cover} and others introduced in Ref.~\cite{crut03}. Here, we will limit
ourselves to the \emph{excess entropy}. The applicability of this analysis to
quantum dynamical systems has been shown in Ref.~\cite{crut06}, where
closed-form expressions are given for some of these quantities when the
generator is known.

We can use the observed behavior, as reflected in the word distribution,
to come to a number of conclusions about how a quantum process generates
randomness and stores and transforms historical information. The
{\em Shannon entropy} of length-$L$ sequences is defined
\begin{align}
\label{eq:HL}
H(L)  &\equiv  - \sum_{ s^L \in {\cal A}^L } \Prob (s^L) \l2 \Prob (s^L) ~.
\end{align}
It measures the average surprise in observing the ``event'' $s^L$.
Ref. \cite{crut03} showed that a stochastic process's informational
properties can be derived systematically by taking derivatives and then
integrals of $H(L)$, as a function of $L$.
For example, the {\em source entropy rate} $\hmu$ is the rate of increase with
respect to $L$ of the Shannon entropy in the large-$L$ limit:
\begin{equation}
    \hmu \equiv \lim_{L \rightarrow \infty} \left[ H(L) - H(L-1) \right] \; ,
\label{ent.def}
\end{equation}
where the units are \emph{bits/measurement} \cite{cover}.  

Ref.~\cite{crut06} showed that the entropy rate of a quantum process can be
calculated directly from its QFG, when the latter is
deterministic. A closed-form expression for the entropy rate in this case is
given by:
\begin{align}
\hmu = - |Q|^{-1} \sum_{i=0}^{|Q|-1}\sum_{j=0}^{|Q|-1}
	|U_{ij}|^2 \log_2 |U_{ij}|^2 ~,
\end{align}
The entropy rate
$\hmu$ quantifies the irreducible randomness in 
processes: the randomness that remains after the correlations and
structures in longer and longer sequences are taken into account.

The latter, in turn, is measured by a complementary quantity.
The amount $I(\past;\future)$ of mutual information \cite{cover} shared
between a process's past $\past$ and its future $\future$ is given by the
\emph{excess entropy} $\EE$ \cite{crut03}. It is the subextensive
part of $H(L)$: 
\begin{equation}
  \EE = \lim_{L \rightarrow \infty} [ H(L) - \hmu L ]\;.
\label{EEfromEntropyGrowth}
\end{equation}
Note that the units here are \emph{bits}.

Ref.~\cite{crut03} gives a closed-form expression for $\EE$ for \emph{order-$R$}
Markov processes---those in which the measurement symbol probabilities depend
only on the previous $R-1$ symbols. In this case,
Eq.~(\ref{EEfromEntropyGrowth}) reduces to:
\begin{align}
\label{eq:eeR}
E = H(R) - R\cdot \hmu~,
\end{align}
where $H(R)$ is a sum over $|\Abet|^R$ terms. Given that the quantum generator
is deterministic we can simply employ the above formula for $\hmu$ and compute
the block entropy at length $R$ to obtain the excess entropy for the
order-$R$ quantum process.

Ref.~\cite{crut06} computes these entropy measures for various example
systems, including the spin-$1$ particle. The results are summarized in
Table~\ref{tab:info}. The value for the excess entropy of the Golden Mean
process obtained by using Eq.~(\ref{eq:eeR}) agrees with the value obtained
from simulation data, shown in Table~\ref{tab:info}. The entropy $\hmu = 2/3$
bits per measurement for both processes, and thus they have the same amount
of irreducible randomness. The excess entropy, though, differs markedly. The
Golden Mean process ($J_y^2$ measured) stores, on average, $\EE \approx 0.25$
bits at any given time step. The Even Process ($J_x^2$ measured) stores, on
average, $\EE \approx 0.90$ bits, which reflects its longer memory of
previous measurements.

\begin{table}[tbp]
\begin{tabular}{|c||c|c|}
\hline
 Quantum & \multicolumn{2}{c|}{Spin-1} \\
Dynamical System        & \multicolumn{2}{c|}{Particle } \\
\hline
Observable & $J_y^2$             & $J_x^2$ \\
\hline
\hline
 $\hmu$ [\emph{bits/measurement}] & ~0.666~	& ~0.666~ \\
 $\EE$   [\emph{bits}] & 0.252 & 0.902 \\
\hline
\end{tabular}
\caption{Information storage and generation for example quantum processes:
  entropy rate $\hmu$ and excess entropy $\EE$.
  }
\label{tab:info}
\end{table}

% **********************************************************************
\section{Conclusion}

We have shown that quantum dynamical systems store information in their
dynamics. The information is accessed via measurement. Closer inspection
would suggest even that information is \emph{created} through measurement.
In any case, the key conclusion is that, since both processes are represented by a
$3$-state QFG constructed from the same internal quantum dynamics, it is the
means of observation alone that affects the amount of memory. 
This was illustrated with the particular examples of the
spin-$1$ particle in a magnetic field. Depending on the choice of observable
the spin-$1$ particle generates different process languages. We showed that
these could be analyzed in terms of the block entropy---a measure of
uncertainty, the entropy rate---a measure of irreducible randomness, and the
excess entropy---a measure of structure. Knowing the (deterministic) QFG
representation, these quantities can be calculated in closed form. 

We established a connection between quantum automata theory and quantum
dynamics, similar to the way \emph{symbolic dynamics} connects classical
dynamics and automata.  
By considering the output sequence of a repeatedly measured quantum system
as a shift system we found quantum processes that are sofic systems. Taking
one quantum system and observing it in one way yields a subshift of finite type.
Observing it in a different way yields a (strictly sofic) subshift of infinite
type. Consequently, not only the amount of memory but also the soficity of a
quantum process depend on the means of observation.

This can be compared to the fact that, classically the Golden Mean and the
Even sofic systems can be transformed into each other by a two-block map. The
adjacency matrix of the
graphs is the same. A similar situation arises here. The unitary matrix, which
is the corresponding adjacency matrix of the quantum graph, is the same for
both processes. The processes can be transformed into each other by expressing
one set of projection operators in the eigenbasis of the other. This
transformation always exists since the operators simply represent different
orthonormal basis sets spanning the Hilbert space.\\

The preceding attempted to forge a link between quantum dynamical systems and
quantum computation by extending concepts from symbolic dynamics to the
quantum setting. We believe the results suggest further study of the properties
of quantum finite-state generators and the processes they generate is necessary
and will shed light on a number of questions in quantum information processing.
One open technical question is whether sofic quantum systems are the closure
of quantum subshifts of finite-type, as they are for classical systems
\cite{weis73}. There are indications that this is not so. For example, as we
noted, quantum finite-state recognizers can recognize nonregular process
languages \cite{wies06b}. 

% **************************** REFERENCES ****************************
\bibliography{ref}
\end{document}